\documentclass[article,twocolumn,prb,showpacs,superscriptaddress]{revtex4-1}

\usepackage{graphicx}
\usepackage{amssymb}
\usepackage{bm}
\usepackage{amsmath}
\usepackage{xfrac}
\usepackage{color}
\usepackage{titlesec}

\usepackage[colorlinks=True,
                 linkcolor=black,
	        citecolor=blue,
	        urlcolor=blue]{hyperref}

\newcommand\3{$_3$}
\renewcommand\a{{\bf a}}
\renewcommand\b{{\bf b}}
\renewcommand\c{{\bf c}}
\renewcommand\d{{\bf d}}
\newcommand\e{{\bf e}}
\newcommand\f{{\bf f}}
\newcommand\g{{\bf g}}
\newcommand\h{{\bf h}}

\newcommand{\pref}[2]{\ref{#1}{\bfseries #2}}

\newcommand{\marvel}{National Centre for Computational Design and Discovery of Novel Materials (MARVEL), \'Ecole Polytechnique F\'ed\'erale de Lausanne, CH-1015 Lausanne, Switzerland}
\newcommand{\dqmp}{Department of Quantum Matter Physics, University of Geneva, 24 Quai Ernest Ansermet, CH-1211 Geneva, Switzerland}
\newcommand{\gap}{Group of Applied Physics, University of Geneva, 24 Quai Ernest Ansermet, CH-1211 Geneva, Switzerland}	
\newcommand{\nims}{National Institute for Materials Science, 1-1 Namiki, Tsukuba, 305-0044, Japan}

\makeatletter
\g@addto@macro\bfseries{\boldmath}
\makeatother

\begin{document}
\title{Determining the phase diagram of atomically thin layered antiferromagnet CrCl\3}
\date{\today}
\author{Zhe Wang}
\affiliation{\dqmp}
\affiliation{\gap}
\author{Marco Gibertini}
\affiliation{\dqmp}
\affiliation{\marvel}
\author{Dumitru Dumcenco}
\affiliation{\dqmp}
\author{Takashi Taniguchi}
\affiliation{\nims}
\author{Kenji Watanabe}
\affiliation{\nims}
\author{Enrico Giannini}
\affiliation{\dqmp}
\author{Alberto F.\ Morpurgo}
\affiliation{\dqmp}
\affiliation{\gap}

\maketitle

{\bf
Changes in the spin configuration of atomically-thin, magnetic van-der-Waals multilayers can cause drastic modifications in their opto-electronic properties. Conversely, the opto-electronic response of these systems provides information about the magnetic state, very difficult to obtain otherwise. Here we show that in CrCl\3 multilayers, the dependence of the tunnelling conductance on applied magnetic field ($H$), temperature ($T$), and number of layers ($N$) tracks the evolution of the magnetic state, enabling the magnetic phase diagram of these systems to be determined experimentally. Besides a high-field spin-flip transition occurring for all thicknesses, the in-plane magnetoconductance exhibits an even-odd effect due to a low-field spin-flop transition. If the layer number $N$ is even, the transition occurs at $\mu_0 H \sim 0$~T due to the very small in-plane magnetic anisotropy, whereas for odd $N$ the net magnetization of the uncompensated layer causes the transition to occur at finite $H$. Through a quantitative analysis of the phenomena, we determine the interlayer exchange coupling as well as the staggered magnetization, and show that in CrCl\3 shape anisotropy dominates. Our results reveal the rich  behaviour of atomically-thin layered antiferromagnets with weak magnetic anisotropy.

}

The recent discovery of magnetism in atomically thin layers exfoliated from bulk van der Waals (vdW) crystals is a major breakthrough~\cite{Lee2016,Wang2016nl,Kuo2016,Lin2016,Du2016,Huang2017,Gong2017,Ghazaryan2018,Deng2018,Fei2018,Wang2018fgt,Bonilla2018,OHara2018}, with important implications in the field of two-dimensional (2D) materials and heterostructures~\cite{Burch_review_2018,Gong_review_2019,Gibertini_review_2019}. The observation of a giant tunnelling magnetoresistance through CrI\3 multilayers~\cite{Klein2018,Song2018,Wang2018,Kim2018}, for instance, makes clear that semiconducting vdW antiferromagnets are very interesting systems~\cite{Jiang2018field,Jiang2018doping,Huang2018}, with potential relevance for technology~\cite{Jiang2019,Song2019} once room-temperature operation is achieved. It is therefore important to identify the microscopic parameters governing the behaviour of these systems, and to understand how they can be determined experimentally. In CrI\3, the observed phenomenology originates from transitions in which the magnetization of individual layers abruptly flips direction under the application of a perpendicular magnetic field~\cite{Klein2018,Song2018,Wang2018,Kim2018} ($H$), as expected when the uniaxial magnetic anisotropy is the dominating energy scale~\cite{Stryjewski1977,Majlis}. The case of weak anisotropy is different, as a richer variety of magnetic transitions is predicted to occur at lower field~\cite{Neel1936,Majlis}, and indeed in bulk vdW layered antiferromagnets the occurrence of spin-flop phases~\cite{Ubbink1953} is well established~\cite{deJongh1974}. However, the phase diagram of weakly anisotropic atomically-thin vdW layered antiferromagnets, where finite-size effects can give rise to new phenomena~\cite{Dieny1990,Nortemann1992,Wang1994,Rossler2004}, has not been determined yet.

Here, we investigate atomically thin multilayers of CrCl\3, a layered vdW antiferromagnetic semiconductor with a small anisotropy that favours spins to lie in the plane of the layers~\cite{Cable1961,Narath1963,Narath1965,Kuhlow1982,McGuire2017} (Fig.~\ref{fig:MG}\a; see Supplementary Notes 1-2 and Figures 1-2). We show that  the extreme sensitivity of the tunnelling current to the magnetic state can be exploited to track the phase diagram of these systems as a function of temperature and magnetic field. The measurements reveal that, down to bilayers, the antiferromagnetic alignment of spins in neighbouring layers is retained, with an easy-plane magnetic anisotropy that can be entirely attributed to magnetostatic effects (i.e. shape anisotropy). In an external magnetic field, a spin-flip transition is observed for all thicknesses and the dependence of the corresponding critical field on the number of layers and field orientation allows us to estimate both the strength of the interlayer exchange coupling as well as the magnitude of the sublattice magnetization. When $H$  is parallel to the layers, we reveal a prominent even/odd effect, with an additional feature that appears at small fields exclusively for  odd $N$, originating from a spin-flop transition associated with the finite Zeeman energy due to the magnetization of one uncompensated layer.

\begin{figure*}
\centering
\includegraphics[width =1\linewidth]{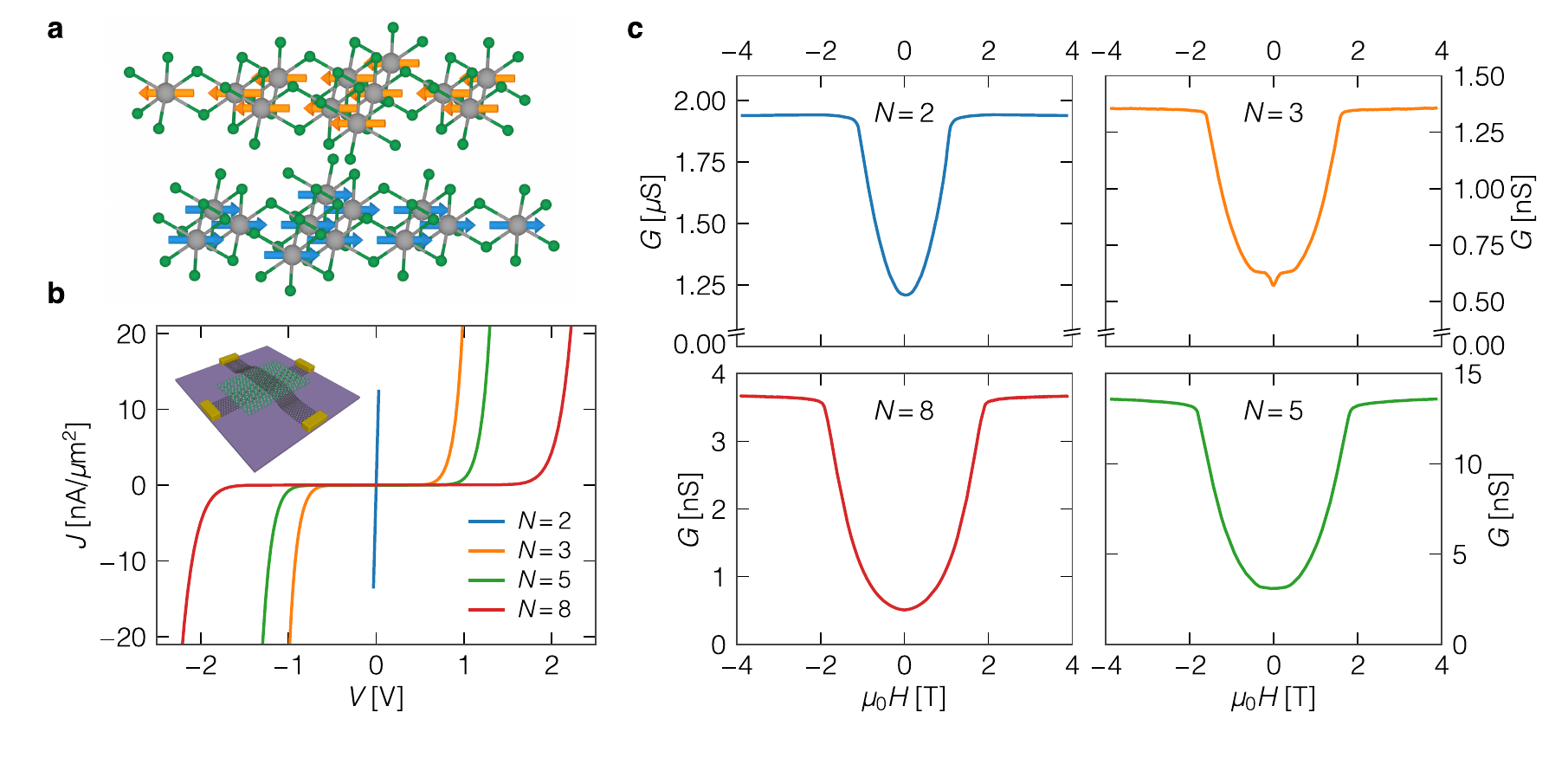}
\caption{\textbf{Tunneling conductance through CrCl\3 multilayers}. {\bf a}: Schematic representation of the crystal structure and of the magnetic moments in adjacent CrCl\3 layers in the absence of an applied magnetic field (grey spheres represent the Cr atoms that are arranged in a honeycomb structure, while Cl atoms are represented in green). At temperatures well below the critical temperature, all spins in a given layer point in the same direction as a consequence of the strong ferromagnetic intralayer exchange coupling, while the weak interlayer  coupling is antiferromagnetic and favours the spins to be antiparallel in neighbouring layers. {\bf b}: Low-temperature $I$-$V$ curves of tunnel barriers formed by CrCl\3 multilayers of different thickness (see value of $N$ in the legend), measured at zero applied magnetic field. Inset: schematic representation of the graphite/ CrCl\3/graphite tunnel junction devices (in actual devices, the tunnel barrier is encapsulated in between a bottom and a top hBN layers to avoid degradation).  {\bf c}: Low-temperature magnetoconductance of CrCl\3 multilayer tunnel barriers, as a function of magnetic field $H$ applied parallel to the layers. The applied voltage is increased upon increasing the device thickness ($V = 10$ mV for $N=2$, $V=0.7$ V for $N=3$, $V=1.1$ V for $N=5$, and $V=1.8$ V for $N=8$), to ensure that the tunneling current is sufficiently large to be measured.}
\label{fig:MG}
\end{figure*}

Our experiments rely on transport measurements done on graphite/CrCl\3/graphite tunnel junctions (see inset of Fig.~\ref{fig:MG}\b\ for a scheme of a device), assembled in a glove box and encapsulated with hexagonal BN to avoid degradation in air (see Supplementary Figs.~3-4 and Methods for details).  At low temperature, all devices show a tunnelling behaviour, in regimes that depend on the multilayer thickness and on the applied voltage. For thin samples, electrons from the graphite electrodes tunnel directly through the barrier, giving rise to a $I$-$V$ curve that is linear at low bias, as shown in Fig.~\pref{fig:MG}{b}. For thicker devices, the current due to direct tunnelling is too small to be detected, and a large bias has to be applied. The resulting $I$-$V$ curves are strongly non-linear, with $\ln(I/V^2)$ scaling proportionally to $1/V$, as characteristic for the Fowler-Nordheim  tunnelling regime~\cite{FN}.

We measure the tunnelling conductance as a function of magnetic field applied parallel to the layers. Sizeable magnetoconductance is observed in all devices (see Fig.~\pref{fig:MG}{c}), with an amplitude reaching a few hundreds percent (depending on the applied bias and thickness of samples), smaller than in CrI\3 but much larger than in non-magnetic tunnel barriers. The conductance $G$ smoothly increases with increasing $H$ and saturates at a value $H_c$, as expected for spins evolving from an antiferromagnetic state at zero field  to a ferromagnetic ordering at saturation. The value of $\mu_0H_c$  depends on the thickness of the sample, and increases from $\sim1.1$~T in bilayers to $\sim1.9$ T in 8-layer and thicker samples. To investigate the occurrence of magnetic transitions we focus our analysis on the evolution of the derivative of the conductance ($dG/dH$) as a function of applied field ($H$) and temperature ($T$). Indeed, $dG/dH$ is sensitive to changes in the relative orientation of the magnetization in the individual layers, and therefore strongly amplifies transitions between different phases.

\begin{figure*}
\centering
\includegraphics[width =1\linewidth]{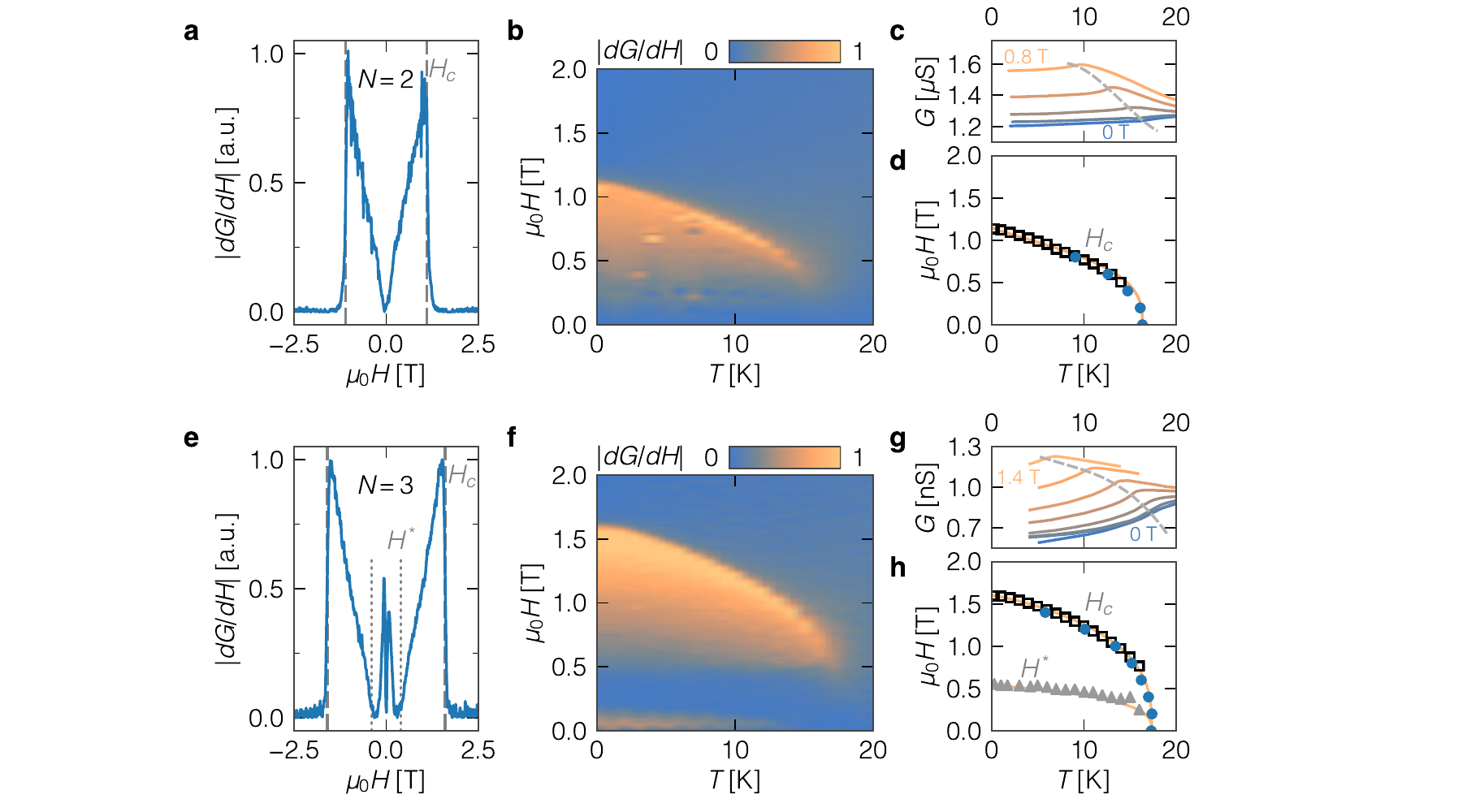}
\caption{\textbf{Phase diagram of bi- and trilayer CrCl\3}. {\bf a}, {\bf e}: magnetic field dependence of $dG/dH$ measured on bi- and trilayer CrCl\3 respectively, at $T=0.3$ K . A sharp jump at $H_c$ is observed in both samples. In addition, in trilayers a minimum in $dG/dH$ is present at finite field $H^*$. {\bf b}, {\bf f}: colour plot of $dG/dH$ as a function of temperature $T$ and applied magnetic field $H$ measured on bi- and trilayer devices. In both devices, the $dG/dH$ jump at $H_c$ shifts to lower fields as $T$ is increased, and finally disappears for $T \simeq  17 $ K, not far from the N\'eel temperature $T_{\rm N}$ of bulk CrCl\3. In the trilayer device in {\bf f}, the blue region below $\mu_0H=0.5$ T tracks the evolution with temperature of the minimum in $dG/dH$ at $H=H^*$.  {\bf c}, {\bf g}: temperature dependence of tunnelling conductance of bi and trilayer devices, measured at different values of applied field $H$ (the field in different curves is increased by 0.2 T per step, starting from $\mu_0H=0$ T). At each value of $H$, a clear feature in the $T$-dependence allows the critical temperature $T_c(H)$ to be determined (taking the derivative $dG/dT$ makes the identification of $T_c(H)$ even simpler; see Supplementary Fig.~7). The dashed line indicates the evolution of the transition temperature. {\bf d}, {\bf h}: magnetic field-temperature ($H$-$T$) phase diagram of --respectively-- bi- and trilayer CrCl\3. The empty black squares corresponds to the position in the $dG/dH$ jump extracted from the measurement of $G$ as a function of $H$, at fixed $T$. The blue dots represent the value of $T_c(H)$ extracted from the measurements of $G$ as a function of $T$ at fixed $H$. As discussed in the main text, these data points trace the phase boundary associated to the spin-flip transition. The triangles in {\bf h} correspond to the value of $H^*$ extracted from $dG/dH$ measurements at fixed $T$, and trace the phase boundary associated to the spin-flop transition (that in odd-$N$ multilayers --such as trilayer CrCl\3-- occurs at finite $H$ despite the vanishing in-plane magnetic anisotropy).}
\label{fig:PD}
\end{figure*}

A sharp jump in $dG/dH$, occurring at an applied magnetic field corresponding to the saturation field $H_c$, is observed in all devices investigated, with Fig.~\pref{fig:PD}{a} and \e\ showing representative data measured on bi- and tri-layer samples at 0.3 K. The colour maps in Fig.~\ref{fig:PD}\b\ and \f\ illustrate the full dependence of $dG/dH$ on $T$ and $H$. The magnetic field at which the jump in $dG/dH$ occurs continuously shifts to smaller values upon increasing $T$ and disappears at a critical N\'eel temperature $T_{\rm N}\simeq 17$ K  (not far from the bulk value, see Supplementary Note 2). The observed evolution resembles the phase boundary of a second-order transition, and is consistent with the temperature and magnetic field dependence of the conductance, where a clear feature at $H_c$ is also present (see Fig.~\pref{fig:PD}\c\ and \g\ and Supplementary Fig.~7). As shown in Fig.~\pref{fig:PD}\d\ and \h\, the data points obtained from either the temperature dependence of $G$ (blue circles) or from the magnetic field dependence of $dG/dH$ (empty squares) are in perfect quantitative agreement (i.e., they trace the same phase boundary). Measurements on trilayers exhibit one additional feature at low field, corresponding to a minimum at finite $H=H^*$, where $dG/dH$ vanishes (Fig.~\pref{fig:PD}{e}). Upon varying $T$, this low-field minimum in $dG/dH$  results in the blue region visible in the colour map of Fig.~\pref{fig:PD}{f}, from which we extract the temperature evolution of $H^*$ (see the triangles in Fig.~\pref{fig:PD}{h}). Similarly to $H_c(T)$, $H^*$ gradually decreases upon increasing $T$, and eventually vanishes at $T \approx T_{\rm N}$, revealing one additional boundary between different magnetic phases.

As we discuss in the rest of the paper, the phase boundaries observed in bi- and trilayer CrCl\3 are representative of the behaviour of all even and odd multilayers. It is therefore useful to understand their origin in terms of the magnetic phases expected to occur in weakly anisotropic layered antiferromagnets~\cite{deJongh1974,deGroot1986}. In such systems, the orientation of the sublattice magnetization is determined by the interplay of three distinct energy scales: the Zeeman energy, the interlayer exchange coupling $J$, and the magnetic anisotropy $K$ (weak anisotropy implies that $K\ll J$, which is indeed known to be the case in bulk CrCl\3). In the bulk, at sufficiently low $H$, the system is in a collinear antiferromagnetic state with the spins pointing along the easy axis, and upon increasing the applied field a spin-flop transition takes place in which the sublattice magnetization aligns in the direction orthogonal to the applied field, with a slight canting. Indeed, in the spin-flop configuration the overall energy balance is favourable above $H_1 \propto \sqrt{JK}$, as the system gains sufficient Zeeman energy (proportional to $H^2/J$) to offset the cost in anisotropy energy (proportional to $K$). A further increase in applied field leads to more canting, until a spin-flip transition eventually occurs (at $H_2 \propto J$), resulting in a state with uniform magnetization aligned with the applied magnetic field. In practice, in CrCl\3, the in-plane magnetic anisotropy is negligible~\cite{Narath1965,Kuhlow1982} (i.e., $K\simeq 0$) and the spin-flop transition occurs at vanishingly small field ($H_1\simeq 0$).
The same is true for even-$N$ multilayers, for which the energy gain of the spin-flop phase over the collinear antiferromagnetic state reads $\Delta E_{\rm even} = - \alpha_N N H^2/J + N K$ (with $\alpha_N$ a positive, $N$-dependent proportionality factor): the first term always dominates since $K\simeq 0$, and the spin-flop transition takes place at $H_1 \simeq 0$. As a result, in even-$N$ multilayers, only one phase boundary corresponding to the spin-flip transition is observed in the experiments.

\begin{figure*}
\centering
\includegraphics{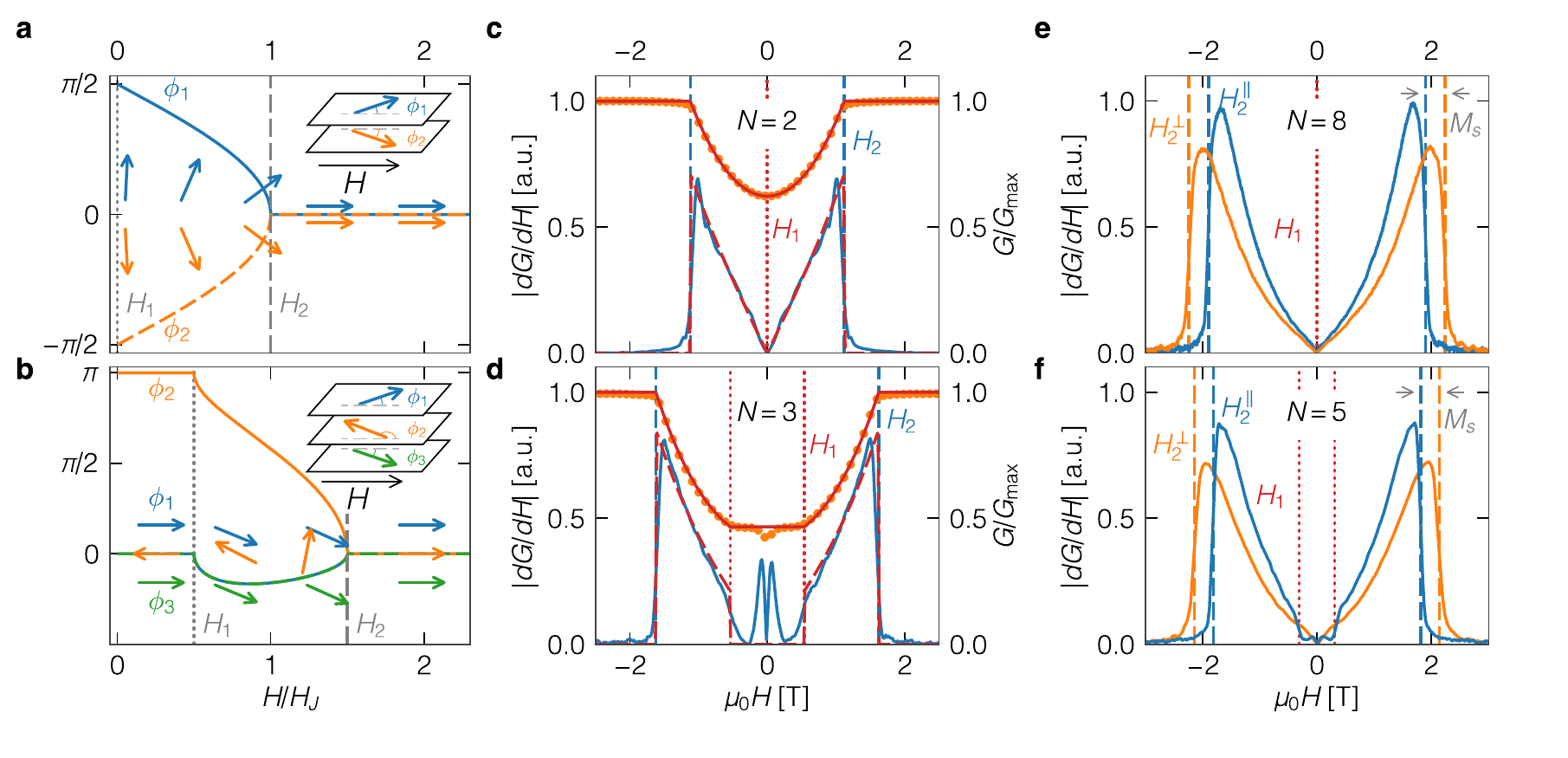}
\caption{\textbf{Even-odd effects in the magnetoconductance of multilayers.} \a, \b: Evolution with applied in-plane magnetic field of the state of bi- (\a)  and trilayer (\b) CrCl\3 (i.e., the orientation of the magnetization in each layer for the case $N=2$ and $N=3$), as obtained from the antiferromagnetic linear-chain model (see Supplementary Note 3). In bilayers the spin-flop transition occurs at zero field, and the canting evolves until the critical spin-flip field $H_2=H_J\equiv 2J/(\mu_0M_s)$ is reached. On the contrary, in trilayers the perfect collinear antiferromagnetic state ($\uparrow \downarrow\uparrow$) is stable up to $H = H_1(=H_J/2)$, at which a spin-flop transition occurs, with the onset of a non-collinear state that evolves until a perfect alignment with the external field is reached at the spin-flip transition at $H=H_2(=3/2H_J)$. \c, \d: Measured (orange circles, same as in Fig.~\pref{fig:MG}{c}) and calculated (red solid line) magnetoconductance $G/G_{\rm max}$ of bi- (\c) and tri-layer (\d) systems. The derivative $dG/dH$ of the experimental data is also reported (blue line) and compared with the theoretical prediction (red dashed line). \e, \f: Measured $dG/dH$ for $N=8$ (\e) and $N=5$ (\f) when the magnetic field is applied parallel (blue line) and perpendicular (orange line) to the layers. Measurements have been performed at same temperature and applied voltage as in Fig.~\pref{fig:MG}{c}. The blue (orange) dashed lines mark the spin-flip field $H_2$ in a parallel (perpendicular) configuration, as obtained from the measured jump in $dG/dH$ (the difference between $H_2^{\perp}$ and $H_2^{\parallel}$ determines the magnetization $M_s$ in each layer). The theoretical prediction for the spin-flop field $H_1$ according to the antiferromagnetic linear-chain model is marked by vertical dotted lines, and is obtained  by setting $H_J$ according to the extracted values of the spin-flip field $H_2$ (vertical red dotted lines) without any additional fitting parameter (in even-$N$ layers, $H_1=0$).
 }
\label{fig:evenodd}
\end{figure*}

The situation is different for odd-$N$ multilayers that possess a net magnetization associated with the presence of an uncompensated layer, giving a finite Zeeman contribution to the total energy. This contribution modifies the energy balance, as $\Delta E_{\rm odd} =  - \alpha_N N H^2/J + N K + \mu_0 M_s H$  ($M_s= 2 g\mu_B S$ is the saturation magnetization per unit cell of a single layer; $S=3/2$ is the spin on Cr atoms). Now, at low $H$ the Zeeman term (linear in $H$) dominates over the energetic gain associated to the spin-flop transition (quadratic in $H$), and forces the magnetization of the uncompensated layer to point parallel to the applied field. It follows that in odd-$N$ multilayers the spin-flop transition occurs at  $H=H_1=JM_s/(\alpha_N N)$, which is finite despite $K \simeq 0$, and can therefore be observed experimentally. Considerations based on the system energetics, therefore, imply that even-$N$ multilayers in our measurements should always exhibit only one observable phase boundary (as $H_1\simeq 0$), whereas two phase boundaries can be observed in odd-$N$ multilayers. They also imply that the field $H_1$ at which the spin-flop transition occurs in odd-$N$ multilayers should decrease upon increasing thickness (approximately as $1/N$), since the effect originates from  the magnetization of one individual uncompensated layer competing against the energetic contribution of all layers.

These predictions can be made fully quantitative using an antiferromagnetic linear-chain model to describe the state of CrCl\3 multilayers, where the magnetization of each layer represents a site in the chain and is coupled to its nearest neighbours through the interlayer exchange energy $J$ (the only unknown parameter in the model;  see Methods Section for details). Considering an in-plane field at low temperature, the state of a generic multilayer --i.e.,  the orientation of the magnetization in each layer-- can be calculated as a function of applied field $H$ by minimising the following magnetic energy per unit cell of the $N$-layer:
\begin{align}
   U_N(\phi_1,\dots,\phi_N;H)  =& J \sum_{i=1}^{N-1} \cos(\phi_{i+1}-\phi_i) \notag\\
   &- \mu_0 H M_s \sum_{i=1}^N \cos(\phi_i)~,
\end{align}
where $\phi_i$ is the angle that the magnetization in the $i$-th layer forms  with the direction of the field.

To understand how the model can be used to interpret our measurements, we start by calculating the linear tunnelling magnetoconductance of bi- and trilayers CrCl\3 and compare it to the experimental results. As a first step, we determine the evolution of the magnetic state (see the arrows and insets in Fig.~\pref{fig:evenodd}{a} and \b\ and Supplementary Note 3). For bilayers (Fig.~\pref{fig:evenodd}{a}), the magnetization in the two layers is nearly perpendicular to the field starting from $H\simeq0$, progressively canting as $H$ is increased, with the spin-flip transition occurring at $H=H_2$. For trilayers (Fig.~\pref{fig:evenodd}{b}), instead, the state does not change for $H$ up to $H_1$, as the magnetization in all layers points in the direction of the applied field (either parallel or antiparallel). Only when $H$ is increased past $H_1$, the orientation of the magnetization in the different layers starts to evolve, and eventually aligns with the applied field at $H=H_2$. Such an evolution of the magnetic state discussed here for bi- and trilayers is characteristic of what happens in all even-$N$ and odd-$N$ multilayers (i.e., the state of even-$N$ multilayers starts changing from $H\simeq 0$, whereas a finite value of $H$ is needed to change the state of odd-$N$ multilayers).

To obtain the tunnelling conductance we assume that each layer acts as a spin filter~\cite{Moodera1988,Worledge2000}, transmitting mainly electrons with their spin parallel to the magnetization (this results in the transmission through adjacent layers to be related to the cosine of the angle between the magnetizations; see Supplementary Note 4). The calculated magnetoconductance $G(H)/G_{max}$ is represented by the continuous lines in Fig.~\pref{fig:evenodd}{c} and \d: the agreement with experimental data is virtually perfect for  bilayers, and excellent for trilayers,  showing that the experimentally observed characteristic fields $H_c$ and $H^*$ can be identified with  $H_2$ and $H_1$. Consistently with the above considerations, in trilayers theory predicts the conductance to be constant for $H<H_1$, and indeed, the measured magnetoconductance curve is rather flat at low field. A small dip is nevertheless present around $H=0$ (whose magnitude is sample~\cite{Klein2019,Cai2019} and bias dependent; see Supplementary Fig.~5) that gives rise to the split peak in $dG/dH$ around $H=0$ (see the blue curve in Fig.~\pref{fig:evenodd}{d}). The origin of this conductance dip, which is present in odd-$N$ multilayers, is currently not clear.

The behaviour of $dG/dH$ discussed for bi- and trilayers is characteristic also for thicker even-$N$ and odd-$N$ multilayers. In all even-$N$ multilayers, $dG/dH$ starts increasing immediately upon the application of a finite magnetic field $H$, and continues to increase up to $H=H_2$, where the spin-flip transition occurs. This  is indeed the behaviour that we observe for $N=8$ (see the blue line in Fig.~\pref{fig:evenodd}{e}), and that was recently reported for $N=4$ in Ref.~\onlinecite{Cai2019}. In odd-$N$ multilayers, instead, $dG/dH$ is predicted to vanish as long as $H<H_1$. In devices where the $H=0$  conductance dip mentioned above is least pronounced, the experimentally observed behaviour is consistent with such expectation, as illustrated by data measured on a $N=5$ (Fig.~\pref{fig:evenodd}{f}) multilayer. Even when the conductance dip at $H=0$ has a larger magnitude and causes $dG/dH$ to peak at small $H$, in odd-$N$ devices a minimum in $dG/dH$ (with $dG/dH\simeq 0$) is still clearly visible at finite $H$ much before the spin-flip transition (see Supplementary Fig.~9), whereas no such a minimum is observed in even-$N$ multilayers. We therefore conclude that the behaviour of $dG/dH$ is distinctly different for even-$N$ and odd-$N$ multilayers or --in other words-- that the experiments indicate the presence of an even-odd effect in the magnetic response of atomically thin CrCl\3 crystals. In odd-$N$ multilayers, we take the value of $H$ at which $dG/dH$ has a minimum  (before starting to monotonously increase as $H$ is swept toward $H_2$) to estimate of the value of $H_1$, the field at which the spin-flop transition occurs (see also supplementary Fig.8).

\begin{figure}
\centering
\includegraphics{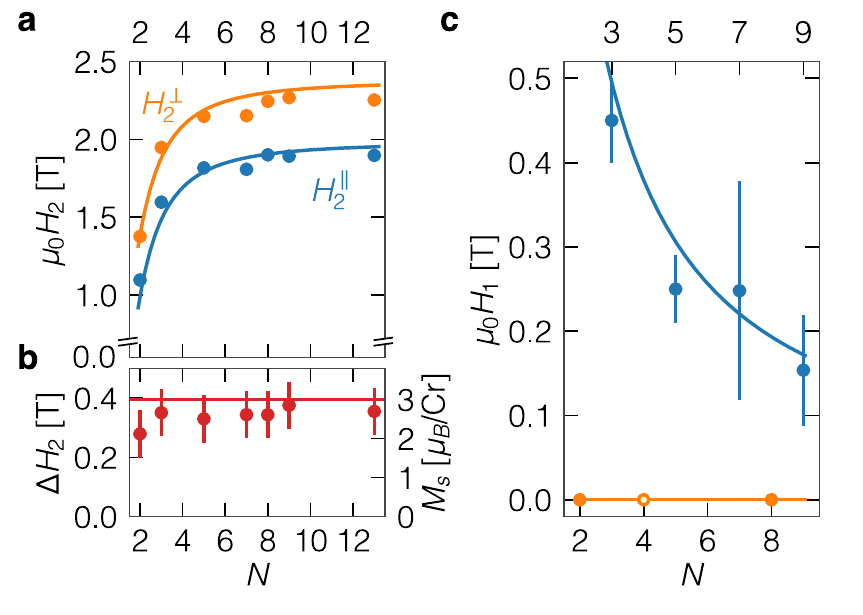}
\caption{\textbf{Thickness evolution of transition fields and saturation magnetization.} \a: Spin-flip field $H_2$ as a function of the number of layers $N$ for magnetic field applied  parallel (blue) and perpendicular (orange) to the layers. Circles represent experimental results extracted from the measured magnetoconductance; solid lines are the theoretical predictions of the antiferromagnetic linear-chain model. We fixed $\mu_0 H_J = 0.99$~T to optimise the agreement between theory and experiments in the parallel configuration; no additional fitting parameter is required to obtain the theoretical curve for the perpendicular case (orange line). \b: Difference $\Delta H_2 = H_2^\perp - H_2^\parallel$ between the measured spin-flip fields plotted in panel \a\, as a function of $N$. The value of $\Delta H_2$ is nearly  constant upon varying $N$, and its value compares well to the expected value of saturation magnetization of the individual layers (right axis), extracted from the known structure of the material (horizontal line). \c: Spin-flop field $H_1$ versus $N$ for even-$N$ (orange) and odd-$N$ (blue) multilayers. The symbols denote data extracted from measurements (the empty orange symbol corresponds to the result in Ref.~\onlinecite{Cai2019}), while solid lines are calculated theoretical predictions obtained using the same parameter $\mu_0 H_J$  as in panel \a\ ,  with no free parameters.  }
\label{fig:thick}
\end{figure}

It is also revealing to compare the evolution of $dG/dH$ measured with the magnetic field applied parallel and perpendicular to the CrCl\3 layers. The orange (blue) curves in Fig.~\pref{fig:evenodd}{e} and \f\ represent data taken with $H$ applied perpendicular (parallel) to the layers (for analogous data taken on multilayers of different thickness see Supplementary Fig.~8). We find that  in all multilayers the spin-flip transition occurs at a higher value of $H$ when the field is applied in the direction perpendicular to the layers (i.e., $H_2$ is systematically larger when $H$ is applied perpendicular to the plane). The value of $H_2$ for all measured multilayers is shown in Fig.~\pref{fig:thick}{a} for field applied in (blue circles, $H_2^\parallel$) and perpendicular to (orange circles, $H_2^\perp$) the plane, and the difference of these two fields, $\Delta H_2 = H_2^\perp - H_2^\parallel$ is shown in Fig.~\pref{fig:thick}{b}. Remarkably, the value of $\Delta H_2$ is nearly exactly the same for all multilayers, independent of $N$, and corresponds quantitatively to the saturation magnetization of an individual CrCl\3 layer (indicated by the horizontal line in Fig.~\pref{fig:thick}{b}; value calculated using the known~\cite{Morosin1964} structure of bulk CrCl\3), similarly to what happens in bulk crystals~\cite{McGuire2017,MacNeill2019}.

The implications of this observation can be understood if we consider that  the spins on the Cr atoms in the material experience an effective ``internal'' field $H_{\rm eff}$, which is the difference between the applied  and the demagnetization fields --i.e., $H_{\rm eff}=H  -\gamma M_v$  ($M_v = M/V$ is the magnetization per unit cell volume $V$ and $\gamma$ the demagnetization factor). In the thin slab geometry of our devices, $\gamma=0$ when the field is applied in-plane, whereas $\gamma=1$ when the field is applied perpendicular. Thus, the observed  difference  $\Delta H_2\simeq M_v$ means that the spin-flip transition occurs at the same effective internal field, irrespective of whether $H$ is applied in-plane or perpendicularly to the plane. This finding is interesting not only because it allows the saturation magnetization of individual CrCl\3 layers to be measured, but also because it implies the absence of significant magnetic anisotropy due to spin-orbit interaction. Indeed, in the presence of magnetocrystalline anisotropy arising from spin-orbit coupling, the value of effective internal field at which the spin-flip transition occurs should depend on direction  (as it happens, for instance, in CrI\3~\cite{Klein2018,Song2018,Wang2018,Kim2018}). We therefore can conclude directly from our measurements that the magnetic anisotropy in atomically thin CrCl\3 multilayers is fully dominated by shape anisotropy.

The thickness dependence of the measured transition fields $H_1$ and $H_2$ is in quantitative agreement with predictions of the antiferromagnetic linear-chain model (see Supplementary Note 5). For the spin-flip transition with the field applied in-plane, the measured values of $H_2$ for different $N$ are represented by the blue circles in Fig.~\pref{fig:thick}{a}. The model predicts the transition to occur at $H^\parallel_2(N) = 2 H_J \cos^2(\frac{\pi}{2N})$ ($H_J=2J/(\mu_0M_s)$ is a field scale associated with the interlayer exchange coupling $J$), with $H_2^\parallel$ doubling when going from the bilayer to the bulk (a behaviour resulting from the reduced antiferromagnetic coupling of the outermost layers that miss one neighbour, which suppresses the overall stiffness of the system to reorient in the direction of the field~\cite{Mills1968}).  This prediction excellently reproduces the experimental points when setting $\mu_0H_J=0.99$~T (see the blue continuous line in Fig.~\pref{fig:thick}{a}). If the field is applied perpendicularly to the layers, the spin-flip transition occurs at $H_2^\perp(N) = H_2^\parallel(N) + M_s/V$ due to the effect of shape anistotropy, a theoretical expression that again matches extremely well the experimental data points (see the orange circles and continuous line in Fig.~\pref{fig:thick}{a}), with the same value of $H_J$ without any additional adjustable parameter.

The same value of $H_J$ also accounts quantitatively for the thickness evolution  of the spin-flop field $H_1$. As shown in Fig.~\pref{fig:thick}{c}, in the even-$N$ multilayers investigated here ($N=2$ and 8) and in Ref.~\onlinecite{Cai2019} ($N=4$), $H_1$ vanishes, due to the negligible in-plane anisotropy (see orange circles in Fig.~\pref{fig:thick}{c}). For odd-$N$ multilayers the measured  $H_1$ values are finite, and decrease with increasing thickness (see blue circles in Fig.~\pref{fig:thick}{c}). The antiferromagnetic linear-chain model predicts that $H_1(N) = H_J \sin(\frac{\pi}{2N})$ (blue continuous line in Fig.~\pref{fig:thick}{c}), in remarkably good agreement with the experimental data with no adjustable parameters, if the value of $H_J$ extracted from the analysis of $H_2$ is used. The antiferromagnetic linear-chain model, therefore, reproduces the full thickness evolution of all measured transition fields ($H_1$, $H_2^\parallel$, and $H_2^\perp$) using as single free parameter the interlayer exchange coupling, for which we extract the value of $J = 86~\mu$eV.

\begin{figure}
\centering
\includegraphics{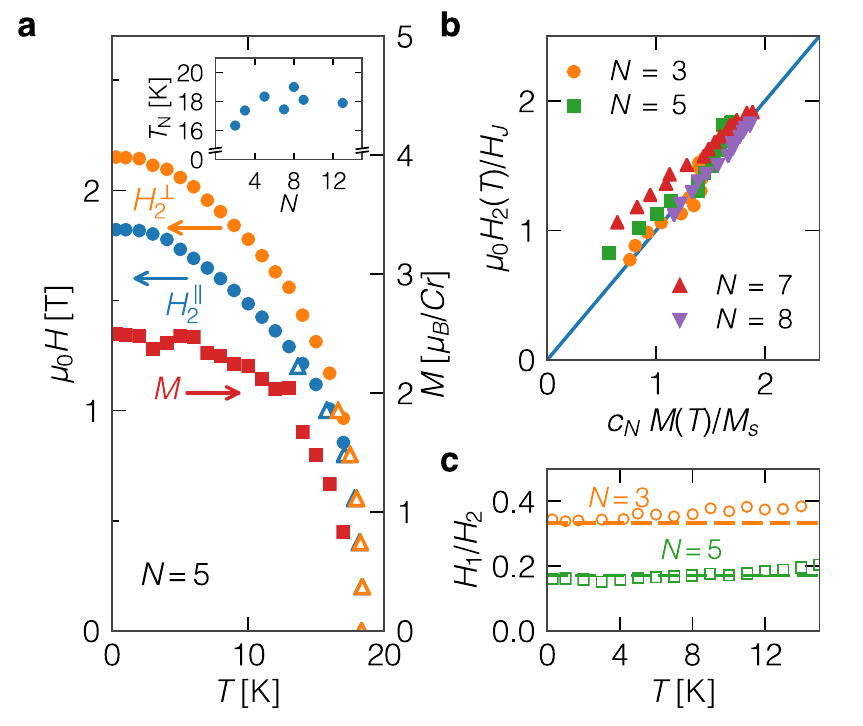}
\caption{\textbf{Temperature dependence.} \a: Temperature evolution of the spin-flip fields $H_2^\parallel$ (blue) and $H_2^\perp$ (orange) for a 5-layer sample.  Solid circles and empty triangles represent points extracted from either $dG/dH$ measurements at fixed $T$ or from $G(T)$ measurements at fixed $H$ (see also Fig.~\pref{fig:PD}{d} and \h). The difference $\Delta H_2$ between the spin-flip fields provides a measure of the layer magnetization $M$ (red squares; see right axis). All quantities ($H_2^\parallel$,  $H_2^\perp$, and $M$) vanish at the same  N\'eel temperature, shown  as a function of the number of layers $N$ in the inset (see Supplementary Note 2 for a comparison with the bulk value).
\b: Correlation between the measured layer magnetization $M$ and spin-flip field $H_2$ measured at different temperatures, for several numbers of layers (the different symbols correspond to the values of $N$ indicated in the legend). The blue straight line provides the theoretical prediction without any fitting parameter. \c: Ratio $H_1/H_2$ between the measured spin-flop and spin-flip fields as a function of temperature for tri- (orange empty circles) and 5-layer (green empty squares) systems. The horizontal dashed line at $1/3$  ($0.171$) represents the temperature-independent theoretical prediction for $N=3$ ($N=5$), again with no fitting parameters. }
\label{fig:temp}
\end{figure}

As a last point in the analysis of the experimental results, we extend our considerations to finite temperature. The dominant role of shape anisotropy enables the $T$-dependence of the layer magnetization to be extracted through the  relationship $\Delta H_2(T) = M(T)/V$. In Fig.~\pref{fig:temp}{a} we show representative results (obtained in the case of a 5-layer sample) for the temperature evolution of the measured spin-flip fields $H_2^\parallel(T)$ and $H_2^\perp(T)$, as well as their difference. All quantities monotonically decrease from their saturation values at low $T$ and  vanish at the same critical temperature $T_{\rm N}~\sim18$~K.  The difference $\Delta H_2$ can be converted into a layer magnetization $M$ as shown on the right axis. It can be proven theoretically (at the level of mean-field theory~\cite{Blazey1971}) that even if $M(T)$ is obtained in the ferromagnetic state at the spin-flip field $H_2$,  this quantity also corresponds to the sublattice magnetization of a single layer in the antiferromagnetic state at zero field. This means that transport measurements allow us to access the sublattice magnetization as a function of temperature, a highly non-trivial conclusion  because it is not obvious how this quantity could be measured otherwise in an atomically thin antiferromagnet.

To show that these considerations are internally consistent and quantitatively in accord within the antiferromagnetic linear-chain model that we have used throughout our analysis, we look at how the relation between $M(T)$ and the spin-flip field $H_2^\parallel(T)$ depends on the multilayer thickness $N$. The model predicts these two quantities to satisfy the proportionality relation  $H^\parallel_2(N,T)/H_J = c_N M(N,T)/M_s $, where $c_N = [1+\cos(\pi/N)]$ is a $N$-dependent constant. All quantities entering this relation are known, as they are either obtained directly from measurements, or can be calculated using the value of $J$ determined from the thickness dependence of $H_2$ (as it is the case of $H_J$). To verify whether the experimental data do obey this relation, in Fig.~\pref{fig:temp}{b} we plot $H^\parallel_2(N,T)/H_J$ as a function of $c_N M(N,T)/M_s$ for several temperatures $T$ and different values of $N$,  for all devices on which we have taken systematic temperature-dependent data in parallel and perpendicular field. We find that indeed all experimental points cluster around the blue, slope-1 line: that is,  as predicted by theory the two quantities do coincide. Finally the model also predicts a proportionality between the spin-flop and spin-flip fields that persists to finite temperature. The expected  relationship $H_1(T)/H_2(T)=\tan\left(\frac{\pi}{2N}\right)/[2\cos\left(\frac{\pi}{2N}\right)]$ is rather well satisfied by the experimental data in Fig.~\pref{fig:temp}{c}.

We conclude that the antiferromagnetic linear-chain model provides an excellent quantitative understanding of the evolution of the magnetic states of the systems investigated here, as a function of applied magnetic field and number of layers, both at low-$T$ and as temperature is varied. This finding is very interesting and somewhat unexpected. Indeed, all theoretical predictions that we invoked in the analysis of the experimental data heavily rely on a mean-field approach, whose validity for 2D magnetic systems without a pronounced easy-axis anisotropy --such as CrCl\3 multilayers-- may be questioned. Possibly, however, what is even more interesting and striking, is the ability that we have demonstrated to experimentally trace the complete phase diagram of a rich class of magnetic systems from rather simple magnetoconductance experiments, at a level that enables the identification of subtle phases originating from finite size effects (such as the spin-flop transition and the associated even-odd effect discussed above, which are absent in bulk crystals). Such a method does  offer important advantages  over other techniques. One --besides its simplicity-- is the possibility to probe extremely small material volumes removing (or drastically limiting) the influence of inhomogeneities and domains, which plague the results of many other experimental techniques that require much larger samples sizes. It is clear that finding suitable techniques to probe the magnetic response of atomically thin magnets represents a urgent need in the field of magnetic 2D materials. The possibility to employ tunnelling magnetoresistance in order to extract a multitude of qualitative and quantitative information about the magnetic properties of different 2D materials discloses extremely promising prospects to sustain progress in this area.

{\it Note:} during the finalization of this manuscript we became aware of additional investigations on the tunnelling magnetoconductance of thin CrCl\3 multilayers~\cite{Klein2019,Kim2019,Cai2019,Kim2019b}, which nonetheless do not explore the systematic thickness and temperature evolution of spin-flip and spin-flop transitions discussed here, nor analyze the data through a systematic quantitative comparison to a theoretical model.

\section*{Acknowledgements}
We sincerely acknowledge Nicolas Ubrig and Hugo Henck for helpful discussions, and  Alexandre Ferreira for technical support. A.F.M. gratefully acknowledges financial support from the Swiss National Science Foundation (Division II) and from the EU Graphene Flagship project. M.G.\ acknowledges support from the Swiss National Science Foundation through the Ambizione program. K.W. and T.T. acknowledge support from the Elemental Strategy Initiative conducted by the MEXT, Japan, A3 Foresight by JSPS and the CREST (JPMJCR15F3), JST.

\section*{Author contributions}
Z.W., M.G., and A.F.M.\ conceived the work. D.D.\ and E.G.\ grew CrCl\3 crystals and performed bulk characterization. T.T.\ and K.W.\ provided high-quality boron nitride crystals. Z.W.\ fabricated all samples and performed all transport measurements. M.G.\ carried out all theoretical modelling. Z.W., M.G., and A.F.M.\ analyzed and interpreted the magnetoconductance data. All authors contributed to writing the manuscript.

\section*{Methods}

\section*{Crystal growth}
CrCl\3 multilayers were exfoliated from bulk crystals that we grew by means of a chemical vapor transport (CVT) method, using commercially available polycrystalline CrCl\3 powder (99.9\%, Alpha Aesar) as starting material. To this end,   the CrCl\3 powder was inserted in a quartz tube (inner diameter 8 mm; the tube length is 13 cm) inside a glove box filled with 99.9999\% Ar.    The tube was subsequently evacuated down to $p \approx 10^{-4}$ mbar, sealed, and inserted  in a horizontal tubular furnace enabling a controlled temperature gradient to be established. The temperatures at the hot and cold ends of the tube were set to be $670^\circ$C and $550^\circ$C, respectively. The tube was left in the furnace under these conditions for seven days, after which the furnace was switched off, letting the tube cool down to room temperature. At the end of this process,  violet, fairly transparent CrCl\3 platelets are found at the cold end of the tube (at the hot end a small amount of greenish powder remains, probably Cr\2O\3 originating from Oxygen present in the commercial starting material). The CrCl\3 thin crystalline platelets have lateral size up to 4mm (limited by the diameter of the tube) and exhibit sharp edges, forming $60$ and $120^\circ$ angles, as expected from their crystal structures (Supplementary Fig.~S1). The structure, stoichiometry, and the magnetic response of the crystals were characterized as discussed in the Supplementary Notes 1-2; the results of this characterization were fully compatible with the properties of CrCl\3 known from the literature.

\subsection*{Sample fabrication}
CrCl\3 multilayers were mechanically exfoliated from the platelet crystals discussed here above. Tunnel junctions consisting of  multilayer graphene/CrCl\3/multilayer graphene were assembled using a by-now common pick-and-lift technique, in a cross geometry illustrated in the inset of Fig.~\pref{fig:MG}{b}. To avoid degradation of thin CrCl\3 multilayers, the exfoliation of CrCl\3 and the heterostructure stacking process were done in a glove box filled with Nitrogen gas, and the whole tunneling junction was encapsulated with hBN before being taken out. Conventional electron beam lithography, reactive-ion etching, electron-beam evaporation (10 nm/50 nm Cr/Ar) and lift-off process were used to contact the multilayer graphene electrodes, away from the CrCl\3 tunnel junction. The thickness of the layers was determined by atomic force microscope measurements performed outside the glove box, on the encapsulated devices, which is possible because of the atomic flatness of all the layers used (including the hBN layer under the CrCl\3 flake; see Supplementary Figs.~3-4).

\subsection*{Transport measurements}
Transport measurements were performed using two different cryostats from Oxford Instruments, a Heliox 3He insert with base temperature $T\simeq 0.25$~K and a cryofree Teslatron system with base temperature $T\simeq 1.5$ K, using home-made low-noise electronics. For measurements performed in the Heliox system, samples were taken out of the insert and manually rotated, to change the angle between magnetic field and CrCl\3. The Teslatron cryostat, instead, is equipped with a sample rotator enabling the devices to be rotated at low-temperature.

\subsection*{Antiferromagnetic linear-chain model}
We introduce here the antiferromagnetic linear-chain model that we use to describe CrCl\3 multilayers. Several variants of this model have appeared in multiple contexts and with different names in the literature (see e.g.\ Ref.~\onlinecite{Rossler2004} and references therein).
The main underlying assumption is that the ferromagnetic intra-layer exchange coupling is so strong (compared to the weak antiferromagnetic inter-layer exchange coupling and the external magnetic field) that each layer behaves as a single unit with uniform magnetization. Within this framework, each layer can be considered as a macroscopic spin which is coupled antiferromagnetically to its neighbours in the adjacent layers. We can thus model any multilayer as a linear chain of macroscopic spins, with an average magnetic energy per unit cell at temperature $T$ given by:
\begin{align}\label{eq:finiteT}
    U_N(T) &= J \sum_{i=1}^{N-1}  \frac{{\bm M}_i(T)\cdot{\bm M}_{i+1}(T)}{M_s^2} \\
    &+ \frac{K_\perp}{2} \sum_{i=1}^N \left(\frac{{\bm M}_i(T)\cdot\hat{\bm z}}{M_s}\right)^2 \notag
    - \mu_0 {\bm H}\cdot \sum_{i=1}^N {\bm M}_i(T)~.
\end{align}
Here $J>0$ is the antiferromagnetic inter-layer exchange coupling (taken for simplicity to be uniform throughout the multilayer), ${\bm M}_i(T)$ is the magnetization per unit cell of the $i$-th layer at temperature $T$, $M_s$ is the saturation magnetization per unit cell of a single layer,  $K_\perp>0$ is the easy-plane anisotropy energy (assuming each layer to lie in the $xy$ plane), and ${\bm H}$ is the external magnetic field. In general, the anisotropy energy has two main contributions, $K_\perp = K_{\rm mc} + K_{\rm sh}$, corresponding to the magnetocrystalline anisotropy ($K_{\rm mc}$, stemming from spin-orbit coupling in the material)  and the shape anisotropy ($K_{\rm sh}$, associated with magnetostatic interactions). As we show in the main text, $K_{\rm mc}$ is neglibible and the magnetic anisotropy is dominated by $K_{\rm sh}=\mu_0 M_s^2/V$ ($V$ being the unit cell volume of CrCl\3). Since $M_s = 2 g\mu_B S$ can be easily computed from the nominal valence state of Cr atoms in CrCl\3 (corresponding to $S=3/2$) and $V$ is available from previous experiments~\cite{Narath1963},  this means that the inter-layer exchange coupling $J$ is the only free parameter in the model.

In the zero-temperature limit, the magnetization in the $i$-th layer can be expressed as ${\bm M}_i = M_s \hat{\bm m}_i$, where $\hat{\bm m}_i$ is a unit vector, and the magnetic energy reads:
\begin{align}\label{eq:mag_ene}
    U_N =& J \sum_{i=1}^{N-1}  \hat{\bm m}_i\cdot\hat{\bm m}_{i+1} + \frac{K_\perp}{2} \sum_{i=1}^N \left(\hat{\bm m}_i\cdot\hat{\bm z}\right)^2  \notag\\
    &- \mu_0 M_s {\bm H}\cdot \sum_{i=1}^N \hat{\bm m}_i~,
\end{align}
which reduces to Eq.~(1) in the main text when the field is applied in-plane ${\bm H} = (H,0,0)$ and we can write $\hat{\bm m}_i = (\cos\phi_i,\sin\phi_i,0)$. The most stable magnetic configuration $\{\hat{\bm m}_i\}_{i=1,\dots,N}$ of any multilayer can be obtained by minimizing the energy $U_N$ for a given value (and orientation) of the applied magnetic field $\bm H$. In Supplementary Note 3 we show in practice how to perform the  minimization of $U_N$ to obtain the magnetic configuration of bi- and trilayers as a function of applied magnetic field.

\section*{Data availability}
All relevant data are available from the corresponding authors
upon reasonable and well-motivated request.

\end{document}